\documentclass[twocolumn,amsmath,amssymb,superscriptaddress]{revtex4}

\usepackage{graphicx}

\usepackage[breaklinks]{hyperref}
\hypersetup{
    bookmarks=false,
    pdfstartview={FitH},
    colorlinks=true,
    linkcolor=blue,
    citecolor=blue,
    urlcolor=blue
}

\begin{document}

\title{Strain dependence of bonding and hybridization across the
metal-insulator transition of VO$_2$}

\author{J.~Laverock}
\affiliation{Department of Physics, Boston University, 590 Commonwealth Avenue,
Boston, Massachussetts, MA 02215, USA}

\author{L.~F.~J.~Piper}
\affiliation{Department of Physics, Boston University, 590 Commonwealth Avenue,
Boston, Massachussetts, MA 02215, USA}
\affiliation{Department of Physics, Applied Physics and Astronomy, Binghamton
University, Binghamton, NY 13902, USA}

\author{A.~R.~H.~Preston}
\author{B.~Chen}
\author{J.~McNulty}
\author{K.~E.~Smith}
\affiliation{Department of Physics, Boston University, 590 Commonwealth Avenue,
Boston, Massachussetts, MA 02215, USA}

\author{S.~Kittiwatanakul}
\affiliation{Department of Physics, University of Virginia,
Charlottesville, VA 22904, USA}
\author{J.~W.~Lu}
\affiliation{Department of Materials Science and Engineering,
University of Virginia, Charlottesville, VA 22904, USA}
\author{S.~A.~Wolf}
\affiliation{Department of Physics, University of Virginia,
Charlottesville, VA 22904, USA}
\affiliation{Department of Materials Science and Engineering,
University of Virginia, Charlottesville, VA 22904, USA}

\author{P.-A.~Glans}
\author{J.-H.~Guo}
\affiliation{Advanced Light Source, Lawrence Berkeley National Laboratory,
Berkeley, California, CA 94720, USA}

\begin{abstract}
Soft x-ray spectroscopy is used to investigate the strain dependence of the
metal-insulator transition of VO$_2$.  Changes in the strength of the V $3d$
- O $2p$ hybridization are observed across the transition, and are linked
to the structural distortion. Furthermore, although the V-V dimerization is
well-described by dynamical mean-field theory, the V-O hybridization is found
to have an unexpectedly strong dependence on strain that is not predicted by
band theory, emphasizing the relevance of the O ion to the physics of VO$_2$.
\end{abstract}

\maketitle

Of the materials that exhibit a metal-insulator transition (MIT),
VO$_2$ has been an exquisite textbook example for the last five decades
\cite{morin1959etc}, with its large ($\sim 10^4$) discontinuity
in the conductivity and the rich tunability of its properties with
alloying or strain. Despite such intense interest, however, the nature
of the transition itself still remains a challenge to explain. In the
past, debate about whether the transition is driven by the lattice
\cite{goodenough1973,carruthers1973} (Peierls physics) or by electron
correlation effects \cite{lederer1972,pouget1974} (Mott-Hubbard physics)
fuelled interest; more recently, the general consensus amongst experiment
and theory alike is for a co-operative model, in which both pictures are
important to the MIT.  Key to understanding such a co-operative model
has been the behavior of the transition with applied strain and alloying
\cite{pouget1974,marezio1972,villeneuve1972,pouget1975}. Technologically,
interest in this material has focused on the dramatic changes of its optical
properties through the transition, coupled with its ultra-fast nature
\cite{cavalleri2001}, that make it an excellent candidate for applications
such as fast optical switches.

The potential application of VO$_2$ as a novel functional material
has recently been accelerated by advances in its thin film growth
\cite{muraoka2002,west2008b}, and such interest has been fuelled by the
possibility of tailoring the MIT (including to below room temperature) through
doping and/or strained thin films epitaxially grown on oriented TiO$_2$
substrates.  However, the introduction of strain to the lattice (at ambient
pressure) raises new questions on the physics of VO$_2$.  In particular,
the role of the lattice has important implications for the timescale of the
transition, and many of the envisaged technological applications of VO$_2$
hinge on its ultra-fast nature.  For example, the structural transition
is known to impose a bottleneck on the timescale of the transition of
bulk VO$_2$ \cite{cavalleri2004}.  Meanwhile, the effects of strain on the
mechanism of the MIT are not well-established: in bulk VO$_2$, small amounts
of applied uniaxial stress stabilize the $M_2$ phase \cite{pouget1975},
whereas for Nb-doped VO$_2$ (of Nb concentrations $\geq 15$\%) the chemical
pressure induces an insulating rutile phase \cite{lederer1972} (i.e.\ with
no accompanying structural distortion).  In this Letter, using a combination
of soft x-ray spectroscopies, we show that the MIT of moderately strained
VO$_2$ still involves the lattice, and reveal that the role of the O ion has
a surprising dependence on the strain that is not anticipated by band theory.

The essential features of the electronic structure of VO$_2$ can be
understood rather well from a simple molecular orbital perspective
\cite{goodenough1973}. In this picture, the V $3d$ orbitals form three
well-separated states in the high-temperature rutile ($R$) structure.
Crucially, the $\pi^*$ (also labeled $e^{\pi}_g$) states overlap in energy
with the so-called $d_{\parallel}$ ($a_{1g}$) states that run along the rutile
$c$-axis ($c_{\rm R}$-axis), leading to a metallic phase. Accompanying the
MIT, a structural distortion into the monoclinic $M_1$ phase leads to the
dimerization of V atoms in the $c_{\rm R}$-axis, splitting the $d_{\parallel}$
state into bonding and anti-bonding states.  Additionally, the tilting of
the VO$_6$ octahedra in the $M_1$ structure increases the V-O hybridization,
and pushes the $\pi^*$ state upwards in energy, deoccupying it and leading
to an insulating phase. However, early measurements of the properties of
Nb- and Cr-doped VO$_2$ \cite{lederer1972,pouget1974} revealed features of
the phase diagram inconsistent with the neglect of electron correlations,
e.g.~an insulating rutile phase and the presence of the $M_2$ phase (in
which only half the V atoms dimerize).  In the model of Zylbersztejn and Mott
\cite{zylbersztejn1975}, strong correlations in the $d_{\parallel}$ band are
screened in the metallic phase by the $\pi^*$ band. In the insulating phase,
these $\pi^*$ states are empty, and the unscreened correlations open the gap.

\begin{figure*}[t!]
\begin{center}
\includegraphics[width=1.00\linewidth,clip]{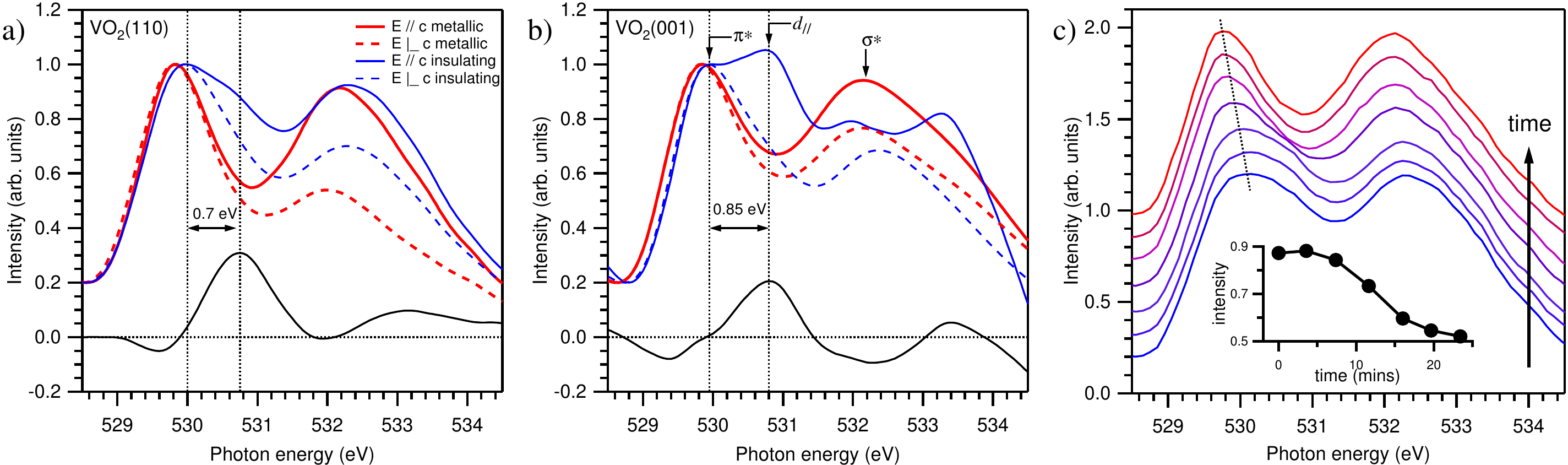}
\end{center}
\vspace*{-0.2in}
\caption{\label{f:xas} (Color online) O $K$-edge XAS spectra for (a)
VO$_2$(110) and (b) VO$_2$(001). Spectra are shown recorded both above
(metallic) and below (insulating) the MIT with incident photons aligned
parallel and perpendicular to the $c_{\rm R}$-axis. At the bottom of
each figure the difference between insulating and metallic spectra are shown.
(c) O $K$-edge spectra recorded during a heating cycle through the transition
for VO$_2$(110). The inset shows the evolution in intensity of the
$d_{\parallel}$ peak.}
\end{figure*}

Attempts to describe the electronic structure from first-principles have had
mixed success: the local density approximation (LDA) has not been able to
account for the insulating phase, leading to a metallic solution for both
$R$ and $M_1$ structures \cite{wentzcovitch1994etc}. On the other hand, the
inclusion of static correlations in the LDA+U method correctly predict an
insulating $M_1$ phase but cannot describe the $R$ phase for reasonable values
of $U$ \cite{korotin2002etc}. More recently, dynamical mean-field theory (DMFT)
calculations \cite{biermann2005,lazarovits2010} and HSE hybrid functional
calculations \cite{eyert2011} have been able to describe both phases well.

X-ray spectroscopy is a powerful tool for addressing the electronic structure
of complex materials, capable of revealing both the unoccupied and occupied
site-specific partial density of states (PDOS) in the form of x-ray absorption
spectroscopy (XAS) and x-ray emission spectroscopy (XES) respectively. Further,
by rotating the polarization vector of the incident x-rays, it is possible to
couple to orbitals of different symmetry.  XAS measurements at the O $K$-edge
of pure, bulk VO$_2$ \cite{abbate1991} have revealed the main features of
the unoccupied O $2p$ PDOS: peaks separated by $\sim 2.5$~eV in the metallic
phase are related to the $\pi^*$ and $\sigma^*$ states. In the insulating
phase, an additional peak owing to the $d_{\parallel}$ orbital develops at
$\sim 1$~eV above $\pi^*$.  More recent polarization-dependent measurements
have unambiguously associated this peak with the $d_{\parallel}$ orbitals,
demonstrating its presence when the polarization vector is parallel to the
$c_{\rm R}$-axis, and absence when it is perpendicular \cite{koethe2006}.

High-quality thin films ($\sim 40$~nm) of VO$_2$ were grown on rutile
TiO$_2$(110) and TiO$_2$(001) oriented substrates by reactive bias target
ion-beam deposition \cite{west2008b}. X-ray diffraction measurements confirm
the epitaxy of VO$_2$ with the substrate, and establish the contracted $c_{\rm
R}$-axis of VO$_2$ grown on TiO$_2$(001) compared with bulk, and expanded
$c_{\rm R}$-axis for VO$_2$ grown on TiO$_2$(110) \cite{xrdnote}.  In the
following, {\em tensilely-strained} VO$_2$/TiO$_2$(110) is referred to as
VO$_2$(110); correspondingly, {\em compressively-strained} VO$_2$/TiO$_2$(001)
is referred to as VO$_2$(001). Soft x-ray spectroscopy measurements were
carried out at beamline X1B of the National Synchrotron Light Source,
Brookhaven and the AXIS endstation of beamline 7.0.1 at the Advanced Light
Source, Berkeley.  XAS measurements were made in total electron yield (TEY)
mode with a beamline energy resolution of 0.2~eV at FWHM, and the photon
energy was calibrated using TiO$_2$ reference spectra of the Ti $L$-edge and
O $K$-edge.  The XES spectra were recorded with a Nordgren-type spectrometer
set to an energy resolution of 0.5~eV at FWHM, and the instrument was
calibrated using a Zn reference spectrum.

In Fig.~\ref{f:xas}, we present O $K$-edge spectra for the two strained samples
both above and below the MIT and for incident photon polarizations parallel
and perpendicular to the $c_{\rm R}$-axis. For the compressively strained
VO$_2$(001) sample, whose $T_{\rm MIT} \approx 300$~K, spectra were recorded in
the metallic $R$ phase at room temperature (RT) and insulating phase at
$\sim 100$~K.  Correspondingly, for the tensilely strained VO$_2$(110), with
$T_{\rm MIT} \approx 345$~K, data were recorded in the $R$ phase at $\sim 400$~K
and insulating phase at RT. For both samples, the spectra are in excellent
qualitative agreement with the data of Refs.~\cite{abbate1991,koethe2006}:
in the metallic phase, two peaks are observed that correspond to the $\pi^*$
and $\sigma^*$ unoccupied states.

Turning our attention to the temperature dependence of the spectra, an
additional peak $\sim 1$~eV above the $\pi^*$ develops in the insulating
phase for $E \parallel c_{\rm R}$. The assignment of this peak as the
$d_{\parallel}$ state that has previously been observed for bulk VO$_2$
\cite{abbate1991,koethe2006} is confirmed by its polarization dependence:
it is absent in both samples for $E \perp c_{\rm R}$. At the
bottom of Fig.~\ref{f:xas}a,b, the difference between the insulating and
metallic spectra for $E \parallel c_{\rm R}$ is shown, highlighting the
contribution from the $d_{\parallel}$ states. For compressive VO$_2$(001),
the peak energy is found to be offset by 0.85~eV from the $\pi^*$ band,
close to the $\sim 1$~eV observed for bulk VO$_2$ \cite{koethe2006}. On the
other hand, for tensilely strained VO$_2$(110), the $d_{\parallel}$ peak
shifts down in energy to 0.7~eV. These results are in good agreement with
the strain-dependence of the $d_{\parallel}$ state from DMFT calculations, in
which the offset between the $\pi^*$ and $d_{\parallel}$ state are calculated
to be 0.7 and 0.9~eV for tensive and compressive $c_{\rm R}$-axis strain
(each of 2\% magnitude) respectively \cite{lazarovits2010}. This observation
of the unoccupied $d_{\parallel}$ band, and its dependence on strain, in the
insulating phase of moderately strained VO$_2$ is evidence of a substantial
V-V dimerizing structural distortion, similar to the $M_1$ or $M_2$ phases
of bulk VO$_2$.
Factor analysis \cite{malinowski1991} of spectra recorded during a heating
cycle through the transition (shown in Fig.~\ref{f:xas}c) revealed only
two eigenvalues. The data were reproduced by a linear combination of the
insulating and metallic end-members, supporting real-space measurements that
suggest the absence of an intermediate phase and lack of dimerization above
the MIT \cite{corr2010}.

\begin{figure}[t!]
\begin{center}
\includegraphics[width=1.0\linewidth,clip]{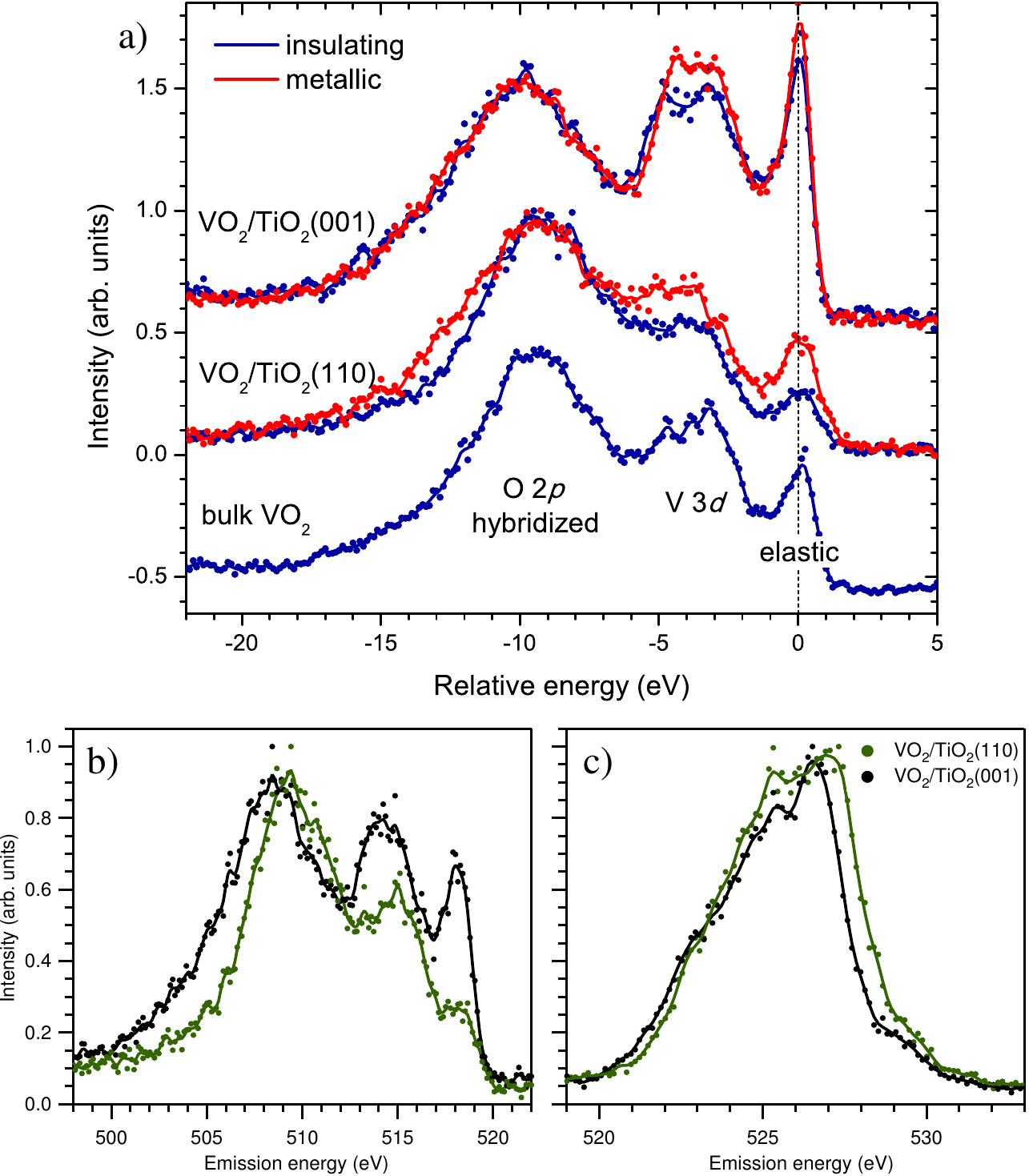}
\end{center}
\vspace*{-0.2in}
\caption{\label{f:xes} (Color online) (a) RXES spectra of VO$_2$(110)
and VO$_2$(001) above and below the MIT recorded at the V $L_3$-edge peak
maximum. The spectrum of bulk $M_1$ VO$_2$ is also shown for comparison.
The spectra have been normalized to the O $2p$ hybridized peak and are
presented on an energy-loss scale and vertically offset. (b) V $L_3$-edge
RXES and (c) O $K$-edge XES spectra of both samples at room temperature on
a common emission energy axis.}
\end{figure}

By resonantly exciting the system at an energy that corresponds to a feature
in the XAS spectrum, it is possible to measure the site-selective occupied
PDOS. In Fig.~\ref{f:xes}a the V $L_3$-edge resonant XES (RXES) spectra
of the two samples in near-grazing geometry
are shown above and below the MIT, and correspond
to the fluorescent decay of occupied V $3d$ states to the empty V $2p$
core hole created in the excitation process (see Ref.~\cite{xrdnote} for details
of the geometry of these measurements).  Three principal features are
observed in these spectra.  Firstly, the peak at 0~eV is due to elastically
scattered x-rays, whose intensity is found to vary with $c_{\rm R}$-axis
strain. Secondly,
the feature centered at $-4$~eV represents a combination of fluorescence
from occupied `pure' V $3d$ orbitals, located just below $E_{\rm F}$,
and low-energy inelastic loss features. Although the separation of these
two components is complicated by their proximity in energy, analysis of
their angular dependence and comparison with photoemission
measurements establish that the loss features are predominantly located below
$-3$~eV. In fact, the strong kink in the VO$_2$(110) spectra at $-3$~eV
separate the two components, analogous to the double-peaked structure
observed for Mo-doped VO$_2$ \cite{schmitt2004}.
Note that the double-peaked structure observed in our data has a different
origin to that observed in photoemission measurements (see, for example,
Refs.~\cite{koethe2006,okazaki2004}).
Finally, the broader
peak centered $-9.5$~eV represents V $3d$ states hybridized with O $2p$
states; the data presented in Fig.~\ref{f:xes}a have been normalized to this
hybridized component. This interpretation is in agreement with other vanadates
\cite{schmitt2004betc}, including Cr-doped VO$_2$ \cite{piper2010}.

Focusing first on the behavior of the spectra across the transition, both
samples exhibit a substantial change in the ratio of the V $3d$ peak to
the O $2p$ hybridized peak, with a stronger contribution from the pure
V $3d$ states in the metallic $R$ phase.  Considering the occupation of
the V $3d$ band to be approximately constant across the transition, this
change in intensity is interpreted as an increase in the hybridization
between O $2p$ and V $3d$ states in going from the metallic to the
insulating phase.  For V$_{1-x}$Cr$_x$O$_2$, a similar change in this
ratio was observed as a consequence of Cr-doping, and was associated with
the decrease in the lattice parameter [in particular, the (110)$_{\rm R}$
lattice spacing] induced by Cr ion substitution \cite{piper2010}.  For bulk
rutile VO$_2$, the nearest-neighbor V-O distance is 1.92~\AA.  However,
in the $M_1$ phase, the structural distortion (dimerization of V-V ions
and particularly their associated twisting in the VO$_6$ octahedra) reduces
this to 1.76~\AA, substantially increasing the overlap of the V $3d$ and O
$2p$ wave functions. Together with our XAS measurements, the RXES data
establish that a structural distortion accompanies the MIT that increases
the overlap between the V $3d$ and O $2p$ wave functions, in a similar manner
to the $M_1$ or $M_2$ bulk phases.

\begin{figure}[t!]
\begin{center}
\includegraphics[width=0.7\linewidth,clip]{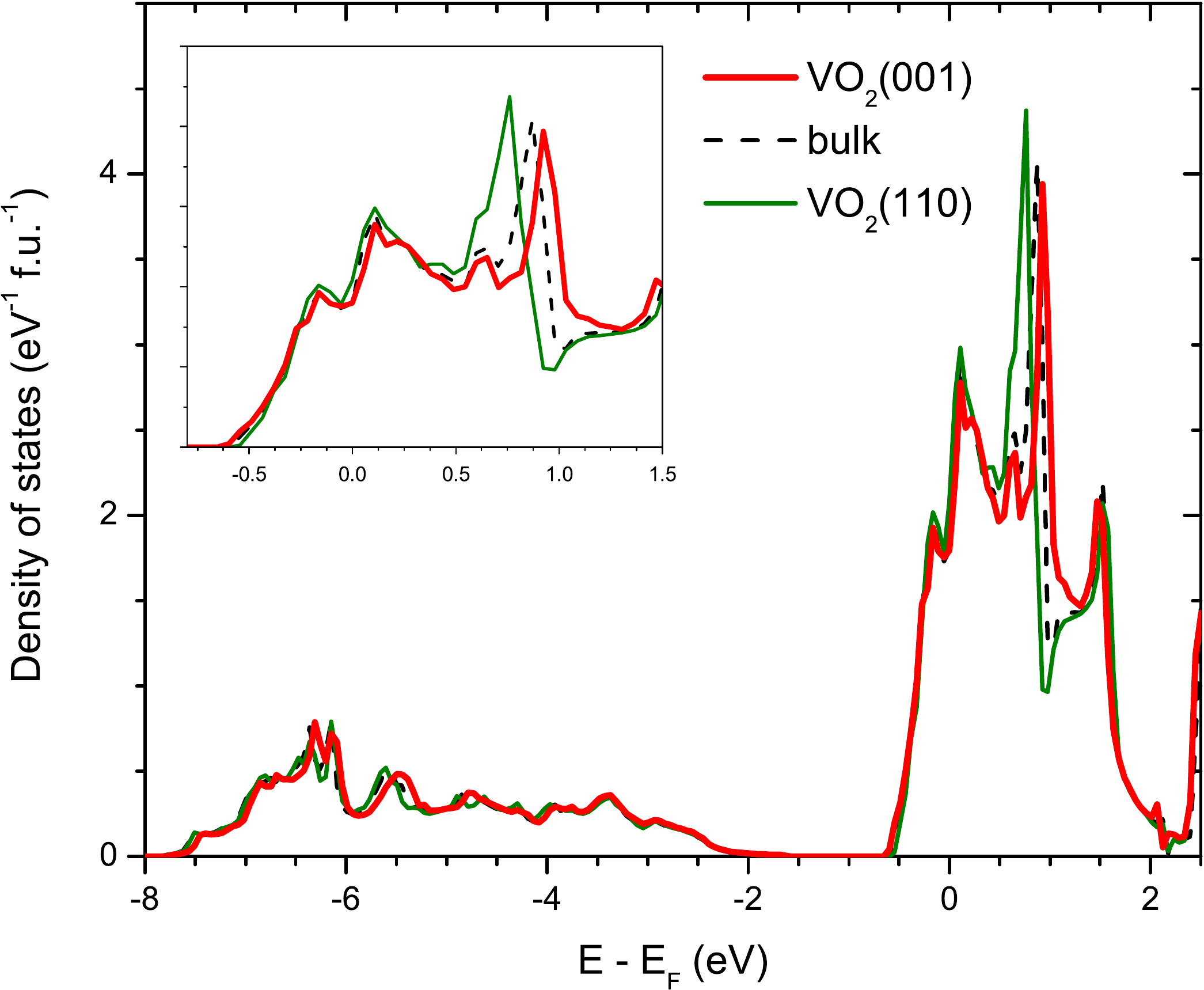}
\end{center}
\vspace*{-0.2in}
\caption{\label{f:vpdos} (Color online) LDA partial V density of states of
metallic rutile VO$_2$ for structures corresponding to the two strained systems
investigated compared with bulk. Shown in the inset is an enlarged view
of the V $3d$ $t_{2g}$ states showing the evolution of the $d_{\parallel}$
peak with strain.}
\end{figure}

Turning now to the behavior of the spectra as a function of strain, it is
evident in Fig.~\ref{f:xes}a that there is a large change in the V $3d$
- O $2p$ hybridization between the two samples. Fig.~\ref{f:xes}b shows
the RXES spectra on a common emission energy axis recorded {\em during the
same measurement}, eliminating ambiguity in the energy calibration of the
spectrometer. Firstly, the location of the O $2p$ (hybridized) feature is
rigidly shifted upwards in energy by almost 1~eV for VO$_2$(110). Owing
to the proximity of loss features, the shift of the V $3d$ PDOS is harder
to accurately judge, but it is clear that any shift of these states is
weaker. This indicates the V and O states are closer together in energy
for VO$_2$(110), facilitating their enhanced hybridization. Indeed, this
enhanced hybridization can be directly visualized in the data through the
relative intensity of the O $2p$ hybridized feature compared with the V $3d$
peak. Qualitatively, such behavior can be understood from the evolution of the
$a_{\rm R}$-axis lattice parameter with strain. For VO$_2$(001), the $a_{\rm
R}$-axis is expanded, accommodating the strain, and the $\pi$-bonded V $3d$
and O $2p$ orbitals are pulled further apart, decreasing their respective
hybridization.  However, such strong strain-dependent changes in the
hybridization between V and O states is not expected from band theory
and hint towards a more significant role for the O ion
than previously considered in the physics of the MIT of VO$_2$.
Shown in Fig.~\ref{f:vpdos} is the V PDOS within the LDA (employing the
FLAPW {\sc Elk} code \cite{elk}) of the metallic phase of both strained and bulk
VO$_2$. The evolution of the $d_{\parallel}$ peak, shown in the inset, is in
agreement with the LDA calculations of Ref.~\cite{lazarovits2010}.  The energy
axis in Fig.~\ref{f:xes}a refers to the excitation energy (rather than
$E_{\rm F}$), and the centroids of the calculated V-O and pure V states agree
reasonably well with experiment with a rigid shift $\sim$~3~eV.  However, the
V-O hybridized states between $-8$ and $-2$~eV show only very weak dependence
on the strain, both in relative intensity and energy, and certainly very
much less than observed in our RXES measurements (Fig.~\ref{f:xes}a).
An accurate model of the electronic
structure of strained VO$_2$ must include such strain induced modifications
to the O states of the electronic structure, not usually directly included
in DMFT calculations.

\begin{figure}[t!]
\begin{center}
\includegraphics[width=0.7\linewidth,clip]{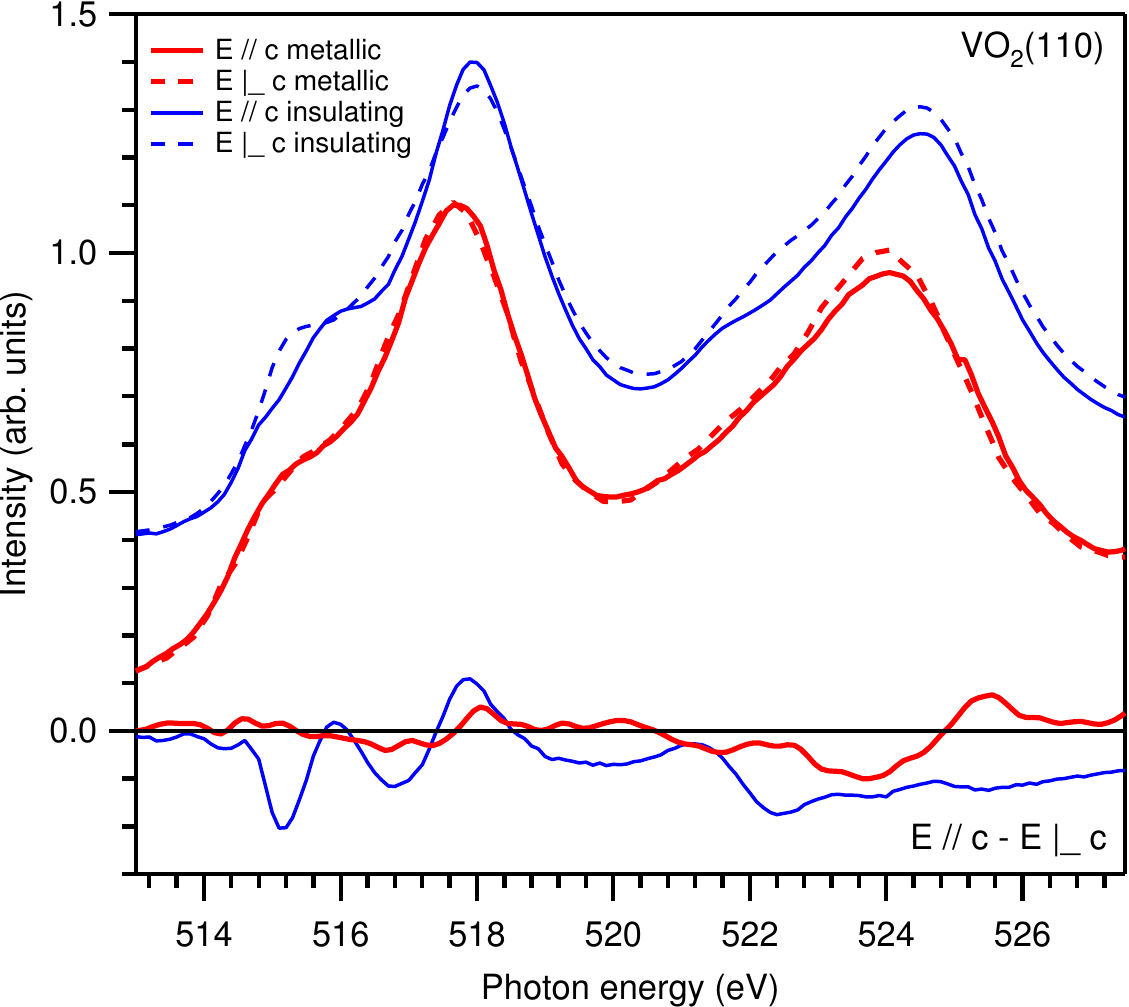}
\end{center}
\vspace*{-0.2in}
\caption{\label{f:vlxas} (Color online) V $L_{3,2}$-edge XAS in both insulating
and metallic phases, shown here for VO$_2$(110). At the bottom of the figure,
the anisotropy of these spectra is shown.}
\end{figure}

O $K$-edge XES spectra are shown for the two samples in Fig.~\ref{f:xes}c
and reflect the change in the local O $2p$ PDOS.
Although there is a slight evolution in the
low-energy onset of the spectra, the dominant difference is the appreciable
narrowing of the bandwidth of VO$_2$(001). Whereas the O $2p$ feature of
the RXES spectra recorded at the V $L_3$-edge are related to the hybridized
portion of the O states, O $K$-edge XES spectra measure the total O $2p$ PDOS.
In particular, the knee observed at $\sim$~529~eV has its origins in the mixing
of the V $3d$ wavefunctions with O $2p$ character.  The broader bandwidth
of VO$_2$(110) is due to the enhanced hybridization with V $3d$ states,
in agreement with the RXES data, and supported by the closer proximity, and
indeed merging, of the knee to the onset of emission from the O $2p$ manifold.

Finally, in Fig.~\ref{f:vlxas} the anisotropy of the V $L_{3,2}$-edge XAS
spectra are shown above and below the MIT of VO$_2$(110); for this system, the
splitting of the $d_{\parallel}$ state is more different (smaller) to that of
bulk VO$_2$ (see Fig.~\ref{f:xas}). Below the transition, these
spectra exhibit strong anisotropy, particularly at the V $L_3$-edge, which
is suppressed in the metallic phase. This anisotropy has been previously
associated with changes in the occupation of the V $3d$ orbitals for bulk
VO$_2$ \cite{haverkort2005}.
It is worth pointing out that electron
correlations were required by Ref.~\cite{haverkort2005} in order to explain
their observations.  These results demonstrate the same orbital switching
occurs for moderately strained VO$_2$, requiring both the monoclinic structural
distortion and the involvement of appreciable electron correlations.

In summary, complementary soft x-ray techniques have been employed to
demonstrate (i) the evolution of V-V dimerization with strain (in excellent
agreement with recent DMFT results), (ii) changes in the hybridization of V-O
states across the MIT (related to the distortion of the VO$_6$ octahedra, and
(iii) the orbital switching that occurs across the MIT. All
of these results are consistent with the distorted monoclinic $M_1$ or $M_2$
low-temperature insulating phase.  Finally, the V-O hybridization is found
to have an unexpectedly strong dependence on strain, indicating a role for
the O ion in the physics of the MIT of VO$_2$ that is often overlooked.
We anticipate that these results will provide a stringent test of theoretical
models of the properties of the MIT in strained VO$_2$.

\section*{Acknowledgements}
We would like to thank S. Sallis for the factor analysis of our data.
The Boston University program is supported in part by the Department of Energy
under Grant No.\ DE-FG02-98ER45680. The ALS, Berkeley, is supported by the
U.S.\ Department of Energy under Contract No.\ DE-AC02-05CH11231. The NSLS,
Brookhaven, is supported by the U.S.\ Department of Energy under Contract
No.\ DE-AC02-98CH10886. SK, JWL, SAW are thankful to the financial support from
the Army Research Office through MURI grant No.\ W911-NF-09-1-0398.


\begin{thebibliography}{99}

\bibitem{morin1959etc}
F.\ J.\ Morin,
\href{http://dx.doi.org/10.1103/PhysRevLett.3.34}
{Phys.\ Rev.\ Lett.\ {\bf 3}, 34 (1959)};
N.\ F.\ Mott,
{\em Metal-Insulator Transitions}, Taylor \& Francis Ltd, London (1974).

\bibitem{goodenough1973}
J.\ B.\ Goodenough and H.\ Y.-P.\ Hong,
\href{http://dx.doi.org/10.1103/PhysRevB.8.1323}
{Phys.\ Rev.\ B {\bf 8}, 1323 (1973)}.

\bibitem{carruthers1973}
E.\ Carruthers and L.\ Kleinman,
\href{http://dx.doi.org/10.1103/PhysRevB.7.3760}
{Phys.\ Rev.\ B {\bf 7}, 3760 (1973)}.

\bibitem{lederer1972}
%P.\ Lederer {\it et al.},
P.\ Lederer, H.\ Launois, J.\ P.\ Pouget, A.\ Casalot and G.\ Villeneuve,
\href{http://dx.doi.org/10.1016/S0022-3697(72)80496-7}
{J.\ Phys.\ Chem.\ Solids {\bf 33}, 1969 (1972)}.

\bibitem{pouget1974}
%J.\ P.\ Pouget {\it et al.},
J.\ P.\ Pouget, H.\ Launois, T.\ M.\ Rice, P.\ Dernier, A.\ Gossard, G.\
Villeneuve and P.\ Hagenmuller,
\href{http://dx.doi.org/10.1103/PhysRevB.10.1801}
{Phys.\ Rev.\ B {\bf 10}, 1801 (1974)}.

\bibitem{marezio1972}
%M.\ Marezio {\it et al.},
M.\ Marezio, D.\ B.\ McWhan, J.\ P.\ Remeika and P.\ D.\ Dernier,
\href{http://dx.doi.org/10.1103/PhysRevB.5.2541}
{Phys.\ Rev.\ B {\bf 5}, 2541 (1972)}.

\bibitem{villeneuve1972}
%G.\ Villeneuve {\it et al.},
G.\ Villeneuve, A.\ Bordet, A.\ Casalot, J.\ P.\ Pouget, H.\ Launois and P.\
Lederer,
\href{http://dx.doi.org/10.1016/S0022-3697(72)80494-3}
{J.\ Phys.\ Chem.\ Solids {\bf 33}, 1953 (1972)}.

\bibitem{pouget1975}
%J.\ P.\ Pouget {\it et al.},
J.\ P.\ Pouget, H.\ Launois, J.\ P.\ D'Haenens, P.\ Merenda and T.\ M.\ Rice,
\href{http://dx.doi.org/10.1103/PhysRevLett.35.873}
{Phys.\ Rev.\ Lett.\ {\bf 35}, 873 (1975)}.

\bibitem{cavalleri2001}
%A.\ Cavalleri {\it et al.},
A.\ Cavalleri, Cs.\ T\'{o}th, C.\ W.\ Siders, J.\ A.\ Squier, F.\ R\'{a}ksi,
P.\ Forget and J.\ C.\ Kieffer,
\href{http://dx.doi.org/10.1103/PhysRevLett.87.237401}
{Phys.\ Rev.\ Lett.\ {\bf 87}, 237401 (2001)}.

\bibitem{muraoka2002}
Y.\ Muraoka and Z.\ Hiroi,
\href{http://dx.doi.org/10.1063/1.1446215}
{Appl.\ Phys.\ Lett.\ {\bf 80}, 583 (2002)}.

\bibitem{west2008b}
%K.\ G.\ West {\it et al.},
K.\ G.\ West, J.\ W.\ Lu, J.\ Yu, D.\ Kirkwood, W.\ Chen, Y.\ H.\ Pei,
J.\ Claassen and S.\ A.\ Wolf,
\href{http://dx.doi.org/10.1116/1.2819268}
{J.\ Vac.\ Sci.\ Technol.\ A {\bf 26}, 133 (2008)}.

\bibitem{cavalleri2004}
%A.\ Cavalleri {\it et al.},
A.\ Cavalleri, Th.\ Dekorsy, H.\ H.\ W.\ Chong, J.\ C.\ Kieffer and R.\ W.\
Schoenlein,
\href{http://dx.doi.org/10.1103/PhysRevB.70.161102}
{Phys.\ Rev.\ B {\bf 70}, 161102(R) (2004)}.

\bibitem{zylbersztejn1975}
A.\ Zylbersztejn and N.\ F.\ Mott,
\href{http://dx.doi.org/10.1103/PhysRevB.11.4383}
{Phys.\ Rev.\ B {\bf 11} 4383 (1975)}.

\bibitem{wentzcovitch1994etc}
R.\ M.\ Wentzcovitch, W.\ W.\ Schulz and P.\ B.\ Allen,
\href{http://dx.doi.org/10.1103/PhysRevLett.72.3389}
{Phys.\ Rev.\ Lett.\ {\bf 72}, 3389 (1994)};
V.\ Eyert,
\href{http://dx.doi.org/10.1002/1521-3889(200210)11:9<650::AID-ANDP650>3.0.CO;2-K}
{Ann.\ Phys.\ (Leipzig) {\bf 11}, 650 (2002)}.

\bibitem{korotin2002etc}
M.\ A.\ Korotin, N.\ A.\ Skorikov and V.\ I.\ Anisimov,
Phys.\ Met.\ Metallogr.\ {\bf 94}, 17 (2002);
%M.\ E.\ Williams {\it et al.},
M.\ E.\ Williams, W.\ H.\ Butler, C.K.\ Mewes, H.\ Sims, M.\ Chshiev and
S.\ K.\ Parker,
\href{http://dx.doi.org/10.1063/1.3072033}
{J.\ Appl.\ Phys.\ {\bf 105}, 07E510 (2009)}.

\bibitem{biermann2005}
%S.\ Biermann {\it et al.},
S.\ Biermann, A.\ Poteryaev, A.\ I.\ Lichtenstein and A.\ Georges,
\href{http://dx.doi.org/10.1103/PhysRevLett.94.026404}
{Phys.\ Rev.\ Lett.\ {\bf 94}, 026404 (2005)}.

\bibitem{lazarovits2010}
%B.\ Lazarovits {\it et al.},
B.\ Lazarovits, K.\ Kim, K.\ Haule and G.\ Kotliar,
\href{http://dx.doi.org/10.1103/PhysRevB.81.115117}
{Phys.\ Rev.\ B {\bf 81}, 115117 (2010)}.

\bibitem{eyert2011}
V.\ Eyert,
\href{http://dx.doi.org/10.1103/PhysRevLett.107.016401}
{Phys.\ Rev.\ Lett.\ {\bf 107}, 016401 (2011)}.

\bibitem{abbate1991}
%M.\ Abbate {\it et al.},
M.\ Abbate, F.\ M.\ F.\ de Groot, J.\ C.\ Fuggle, Y.\ J.\ Ma, C.\ T.\ Chen, F.\
Sette, A.\ Fujimori, Y.\ Ueda and K.\ Kosuge,
\href{http://dx.doi.org/10.1103/PhysRevB.43.7263}
{Phys.\ Rev.\ B {\bf 43}, 7263 (1991)}.

\bibitem{koethe2006}
%T.\ C.\ Koethe {\it et al.},
T.\ C.\ Koethe, Z.\ Hu, M.\ W.\ Haverkort, C.\ Sch\"{u}{\ss}ler-Langeheine, F.\
Venturini, N.\ B.\ Brookes, O.\ Tjernberg, W.\ Reichelt, H.\ H.\ Hsieh, H.-J.\
Lin, C.\ T.\ Chen and L.\ H.\ Tjeng,
\href{http://dx.doi.org/10.1103/PhysRevLett.97.116402}
{Phys.\ Rev.\ Lett.\ {\bf 97}, 116402 (2006)}.

\bibitem{xrdnote}
See Supplemental Material for details of the sample growth and
characterization, and of the anisotropy of the V $L_3$-edge RXES measurements.

\bibitem{malinowski1991}
E.\ R.\ Malinowski,
{\em Factor Analysis in Chemistry}, Wiley, New York (1991).

\bibitem{corr2010}
%S.\ A.\ Corr {\it et al.},
S.\ A.\ Corr, D.\ P.\ Shoemaker, B.\ C.\ Melot and R.\ Seshadri,
\href{http://dx.doi.org/10.1103/PhysRevLett.105.056404}
{Phys.\ Rev.\ Lett.\ {\bf 105}, 056404 (2010)}.

\bibitem{schmitt2004}
%T.\ Schmitt {\it et al.},
T.\ Schmitt, L.-C.\ Duda, M.\ Matsubara, A.\ Augustsson, F.\ Trif, J.-H.\ Guo,
L.\ Gridneva, T.\ Uozumi, A.\ Kotani and J.\ Nordgren,
\href{http://dx.doi.org/10.1016/S0925-8388(03)00575-9}
{J.\ Alloy Compd.\ {\bf 362}, 143 (2004)}.

\bibitem{okazaki2004}
%K.\ Okazaki {\it et al.},
K.\ Okazaki, H.\ Wadati, A.\ Fujimori, M.\ Onoda, Y.\ Muraoka and Z.\ Hiroi,
\href{http://dx.doi.org/10.1103/PhysRevB.69.165104}
{Phys.\ Rev.\ B {\bf 69}, 165104 (2004)}.

\bibitem{schmitt2004betc}
%T.\ Schmitt {\it et al.},
T.\ Schmitt, L.-C.\ Duda, M.\ Matsubara, M.\ Mattesini, M.\ Klemm,
A.\ Augustsson, J.-H.\ Guo, T.\ Uozumi, S.\ Horn, R.\ Ahuja, A.\ Kotani and
J. Nordgren,
\href{http://dx.doi.org/10.1103/PhysRevB.69.125103}
{Phys.\ Rev.\ B {\bf 69}, 125103 (2004)};
%T.\ Schmitt {\it et al.},
T.\ Schmitt, A.\ Augustsson, J.\ Nordgren, L.-C.\ Duda, J.\ H\"{o}wing,
T.\ Gustafsson, U.\ Schwingenschl\"{o}gl and V.\ Eyert,
\href{http://dx.doi.org/10.1063/1.1861125}
{Appl.\ Phys.\ Lett.\ {\bf 86}, 064101 (2005)}.

\bibitem{piper2010}
%L.\ F.\ J.\ Piper {\it et al.},
L.\ F.\ J.\ Piper, A.\ DeMasi, S.\ W.\ Cho, A.\ R.\ H.\ Preston, J.\ Laverock,
K.\ E.\ Smith, K.\ G.\ West, J.\ W.\ Lu and S.\ A.\ Wolf,
\href{http://dx.doi.org/10.1103/PhysRevB.82.235103}
{Phys.\ Rev.\ B {\bf 82}, 235103 (2010)}.

\bibitem{elk}
%J.\ K.\ Dewhurst {\it et al.},
J.\ K.\ Dewhurst, S.\ Sharma, L.\ Nordst\"{o}m, F.\ Cricchio and F.\ Bultmark,
\href{http://elk.sourceforge.net}
{\tt{http://elk.sourceforge.net}} (2009).

\bibitem{haverkort2005}
%M.\ W.\ Haverkort {\it et al.},
M.\ W.\ Haverkort, Z.\ Hu, A.\ Tanaka, W.\ Reichelt, S.\ V.\ Streltsov, M.\ A.\
Korotin, V.\ I.\ Anisimov, H.\ H.\ Hsieh, H.-J.\ Lin, C.\ T.\ Chen, D.\ I.\
Khomskii and L.\ H.\ Tjeng,
\href{http://dx.doi.org/10.1103/PhysRevLett.95.196404}
{Phys.\ Rev.\ Lett.\ {\bf 95}, 196404 (2005)}.

\end{thebibliography}
\end{document}